\documentclass[twocolumn,prb,showpacs,multicol,preprintnumbers,amsmath,amssymb]{revtex4}
\usepackage{graphicx}
\begin{document}
\title{Effect of  point-contact transparency on coherent mixing of Josephson and
transport supercurrents}
\author{Gholamreza Rashedi$^{1}$, Yuri A. Kolesnichenko$^{1,2}$}
\address{$^1$ Institute for Advanced Studies in Basic Sciences, 45195-159,
Zanjan, Iran\\ $^2$ B.Verkin Institute for Low Temperature Physics
 Engineering of National Academy of Sciences of Ukraine, 47,
  Lenin ave , 61103, Kharkov, Ukraine}
\date{\today}
\begin{abstract}
The influence of electron reflection on dc Josephson effect in a
ballistic point contact with transport current in the banks is
considered theoretically. The effect of finite transparency on the
vortex-like currents near the contact and at the phase difference
$\phi =\pi ,$ which has been predicted recently \cite{KOSh}, is
investigated. We show that at low temperatures even a small
reflection on the contact destroys the mentioned vortex-like
current states, which can be restored by increasing of the
temperature.
\pacs{ 74.50.+r, 73.23.Ad, 74.78.Db}
\end{abstract}
\maketitle
\section{Introduction}
The investigations of Josephson effect manifestations in different
systems are continuing because of it's importance both for basic
science and industry. A point contact between two massive
superconductors (S-c-S junction) is one of the possible Josephson
weak links. A microscopic theory of the stationary Josephson
effect in ballistic point contacts between conventional
superconductors was developed in \cite{KO}. Later, this theory was
generalized for a pinhole model in $^{3}He$ \cite{Kurk,Yip}, for
point contacts between ''$d$-wave'' \cite{AmOmZ,Yip1}, and triplet
superconductors \cite{MShK}. The Josephson effect is the phase
sensitive instrument for the analysis of an order parameter in
novel (unconventional) superconductors, where current-phase
dependencies $I_{J}\left( \phi \right) $ may differ essentially
from those in conventional superconductors \cite
{AmOmZ,Yip1,MShK}. In some cases the model with total transparency
of the point contact does not quite adequately correspond to the
experiment, and the electron reflection should be taken into
account. The influence of electron reflection on the Josephson
current in ballistic point contacts was first considered by
Zaitsev \cite{Zaitsev}. He had shown that reflection from the
contact not only changes the critical value of current, but also
the current-phase dependence $I_{J}\left( \phi \right) \sim \sin
\left( \phi /2\right) $ at low temperature which has been
predicted in \cite{KO}. The current-phase dependence for small
values of transparency, $D\ll 1$, is transformed to the
$I_{J}\left( \phi \right) \sim \sin \phi $, similar to the planar
tunnel junction. The effect of transparency for point contact
between unconventional (d-wave) superconductors is studied in the
papers \cite{Amin,Barash,Shumeiko}. The non-locality of Josephson
current in point contacts was investigated in \cite{Heida}. The
authors of \cite{Heida} concentrated on the influence of magnetic
field on the zero voltage supercurrent through the junction. They
found an periodic behavior in terms of magnetic flux and
demonstrated that this anomalous behavior is a result of a
non-locality supercurrent in the junction. This observation was
explained theoretically in \cite{Zagoskin}. Recently an influence
of transport supercurrent, which flows in the contacted banks and
is parallel to the interface, to the Josephson effect in point
contacts has been analyzed theoretically\cite{KOSh}. It was found
that a non-local mixing of two superconducting currents results in
the appearance of two vortex-like current states in vicinity of
the contact, when the external phase difference is $\phi \sim \pi
$. The Josephson current through superconducting weak link is a
result of quantum interference between order
parameters with phase difference $\phi $. Obviously, the finite reflection $%
R=1-D$ of electrons from the Josephson junction suppresses this
interference and it must influence the vortex-like current states,
which are predicted in \cite{KOSh}. In this paper we study the
effect of finite transparency on the current-phase dependence and
distribution of the superconducting current near the ballistic
point contact in the presence of homogeneous current states far
from the contact. We show that at low temperatures $\left(
T\rightarrow 0\right) $ the electron reflection destroys the
mentioned vortex-like current states even for a very small value
of reflection coefficient $R\ll 1.$ On the other hand we have
found that, as the temperature increases the vortices are restored
and they exist for transparency as low as $D=\frac{1}{2}$ in the
limit of $T\rightarrow T_{c}$. The organization of the rest of the
paper is as follows. In Sec.$\text{II}$ we describe the model of
the point contact, quasiclassical equations for Green's functions
and boundary conditions. The analytical formulas for the Green
functions are derived for a ballistic point contact with arbitrary
transparency. In Sec.$\text{III}$ we apply them to analyze a
current state in the ballistic point contact. The influence of the
transport current on the Josephson current and vice versa at the
contact plane is considered. In Sec. $\text{IV}$ we present the
numerical results for the distribution of the current in the
vicinity of the contact. We end in sec.$\text{V}$ with some
conclusions.
\section{Formalism and Basic Equations}
\begin{figure}
 \includegraphics[width=\columnwidth]{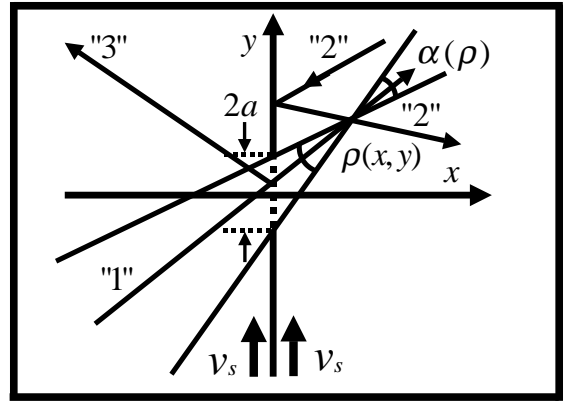}\caption{\label{fig1}
Model of the contact as a slit in the thin insulating partition.}
\end{figure}

We consider the Josephson weak link as a microbridge between thin
superconducting films of thickness $d$. The length $L$ and width
$2a$ of the microbridge, are assumed to be less than the coherence
length $\xi _{0}$. On the other hand, we assume that $L$ and $2a$
are much larger than the Fermi wavelength $\lambda _{F}$ and use
the quasiclassical approach. There is a potential barrier in the
contact, resulting in a finite probability for the electron that
is to be reflected back. In the banks of
superconductors a homogeneous current with a superconducting velocity $%
\bf{{v}_{s}}$ flows parallel to the partition. We choose the
$y$-axis along $\bf{{v}_{s}}$ and the $x$-axis perpendicular to
the boundary; $x=0 $ is the boundary plane (see Fig.\ref{fig1}).
If the film thickness $d\ll \xi _{0}$ then in the main
approximation in terms of the parameter $d/\xi _{0}$ the
superconducting current depends on the coordinates in the plane of the film $%
{\bf \rho} =(x,y)$ only. The superconducting current in the
quasiclassical approximation
\begin{equation}
{\bf j}({\bf \rho} ,{\bf v}_{s})=-2\pi ieN(0)T\sum_{\omega
_{n}}\left\langle {\bf v}_{F}g({\bf v}_{F},{\bf \rho},  {\bf
v}_{s}) \right\rangle _{{\bf v}_{F}} \label{Current}
\end{equation}
is defined by the energy integrated Green's function
\begin{equation}
\widehat{G}=\widehat{G}(\omega _{n},{\bf v}_{F},{\bf \rho}, {\bf
v}_{s})=\left(
\begin{array}{cc}
g & f \\
f^{+} & -g
\end{array}
\right),   \label{Green_fun}
\end{equation}
which in the ballistic case satisfies the Eilenberger equation of
the form \cite{Eilen,Belzig}
\begin{equation}
{\bf v}_{F}\cdot \frac{\partial}{\partial {\bf \rho }}\widehat{G}+
\left[ \widetilde{\omega }\widehat{\tau }_{3}+\widehat{\Delta
},\widehat{G} \right] =0, \label{Eilenberger Eq}
\end{equation}
with normalization condition, $g^{2}+ff^{\dagger }=1$. Here $N(0)$
is the density of states at the Fermi level, $\widetilde{\omega
}=\omega _{n}+i{\bf p}_{F}\cdot {\bf v}_{s}$, ${\bf v}_{F}$ and
${\bf p}_{F}$ are the electron velocity and momentum on the Fermi
surface, $\omega _{n}=(2n+1)\pi T$ are the Matsubara frequencies,
$n$ is an integer number, $ {\bf v}_{s}$ is the superfluid
velocity and $T$ is the temperature, $ \widehat{\Delta }=\left(
\begin{array}{cc}
0 & \Delta  \\
\Delta ^{\ast } & 0
\end{array}
\right) $ is the order parameter matrix, $\widehat{\tau }_{3}$ is
the Pauli matrix. Eqs.(\ref{Eilenberger Eq}) should be
supplemented by the equation for the superconducting order
parameter $\Delta $
\begin{equation}
\Delta ({\bf \rho},{\bf v}_{s},T)=2\pi \lambda T\sum_{\omega
_{n}>0}\left\langle f({\bf v}_{F},{\bf \rho} ,{\bf
v}_{s})\right\rangle _{ {\bf v}_{F}}  \label{Eq for Delta}
\end{equation}
where $\lambda $ is the constant of pairing interaction and
$\left\langle ...\right\rangle _{{\bf v}_{F}}$ is the averaging
over directions of $ {\bf v}_{F}$. As it was shown in \cite{KO} in
the zero approximation in terms of the small parameter $a/\xi
_{0}\ll 1$ for a self-consistent solution of the problem it is not
necessary to consider Eq.(\ref{Eq for Delta}). The model, in which
the order parameter is constant in the two half-spaces $\Delta
({\bf \rho},{\bf v}_{s},T)=\Delta ({\bf v} _{s},T)\exp
($sgn$\left( x\right) \frac{i\phi }{2})$ ($\phi $ is the phase
difference between superconductors), can be used.
\begin{figure}
 \includegraphics[width=\columnwidth]{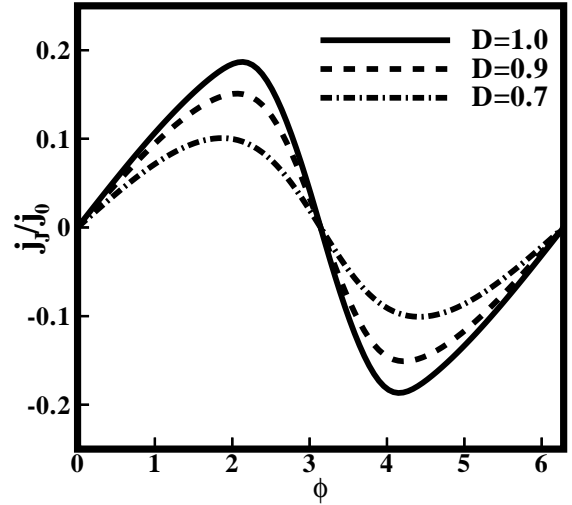}\caption{\label{fig2}Josephson
current $j_{J}$ versus phase $\protect\phi$ for $T/T_c=0.1$,
$q=0.5$ and $j_{0}=4\protect\pi \left| e\right| N(0)v_{F}T_{c}$.}
\end{figure}
In the same approximation the velocity ${\bf v}_{s}$ does not
depend on the coordinates. The Eq.(\ref{Eq for Delta}) enables us
to calculate a spatial distribution of the order parameter $\Delta
({\bf \rho} )$ in the next order approximation in terms of the
parameter $a/\xi _{0}.$ Solutions of Eqs.(\ref{Eilenberger Eq})
should satisfy Zaitsev's boundary conditions (\cite{Zaitsev})
across the
contact $x=0,|y|\leq a$ and specular reflection condition for $x=0,|y|\geq a$%
. In addition, far from the contact, solutions should coincide
with the bulk solutions. The Zaitsev boundary conditions at the
contact can be written as \cite{Zaitsev,Amin,Barash,Belzig}
\begin{equation}
\widehat{d}^{~l}=\widehat{d}^{~r}\equiv \widehat{d}
\label{Zaitsev1}
\end{equation}
\begin{equation}
{\frac{D}{2-D}}\left[
(1+{\frac{\widehat{d}}{2}})\widehat{s}^{~r},\widehat{s}
^{~l}\right] =\widehat{d}\ \widehat{s}^{~l2}  \label{Zaitsev2}
\end{equation}
where
\begin{equation}
\widehat{s}^{~r}=\widehat{G}_{\omega }^{r}({\bf
v}_{F},x=0)+\widehat{G} _{\omega }^{r}({{\bf v}_{F}}^{\prime},x=0)
\end{equation}
\begin{equation}
\widehat{d}^{~r}=\widehat{G}_{\omega }^{r}({\bf
v}_{F},x=0)-\widehat{G} _{\omega }^{r}({{\bf v}_{F}}^{\prime
},x=0)
\end{equation}
with ${{\bf v}_{F}}^{\prime }$ being the reflection of ${\bf
v}_{F}$ with respect to the boundary and $D$ is the transparency
coefficient of point contact. Indexes $l$ and $r$ denote that the
Green function are taken at the left $\left( x=-0\right) $ or
right $\left( x=+0\right) $ hand from
the barrier.\ Similar relations also hold for $\widehat{s}^{~l}$ and $%
\widehat{d}^{~l}$. In general, $D$ can be momentum dependent. For
simplicity in our calculations we assumed that $D$ is independent
of the Fermi velocity direction.\

\section{Current-phase dependencies for Josephson and tangential currents.}

Making use of the solution of Eilenberger equations
(\ref{Eilenberger Eq}), we obtain the following expression for the
current density (\ref{Current}) at the slit:\begin{equation} {\bf
j}_{cont}={\bf j}(x=0,\left| y\right| <a,\phi ,{\bf v}_{s})=
\label{j1}
\end{equation}
$$4\pi eN(0)v_{F}T\sum\limits_{\omega >0}\left\langle \widehat{{\bf
v}} \text{Im}\left( \frac{\widetilde{\omega }\Omega -i\eta D\Delta
^{2}\sin \frac{\phi }{2}\cos \frac{\phi }{2}}{\Omega ^{2}-\Delta
^{2}D{(\sin \frac{ \phi }{2})}^{2}}\right) \right\rangle
_{\widehat{{\bf v}}}$$ where, $\Omega =\sqrt{\widetilde{\omega
}^{2}+\Delta ^{2}}$, $\widehat{ {\bf v}}={\bf v}_{F}/v_{F}$ is the
unit vector and $\eta =\text{sgn} (v_{x})$. We should require
$\text{Re}\Omega>0,$ which fixes the sign of the square root to be
$\text{sgn}({\bf p}_{F}{\bf v}_{s})$. In the case, ${\bf
v}_{s}\neq 0$, the current (\ref{j1}) has both ${\bf j}_{J} $ and
${\bf j}_{y}$ components.
\begin{figure}
 \includegraphics[width=\columnwidth]{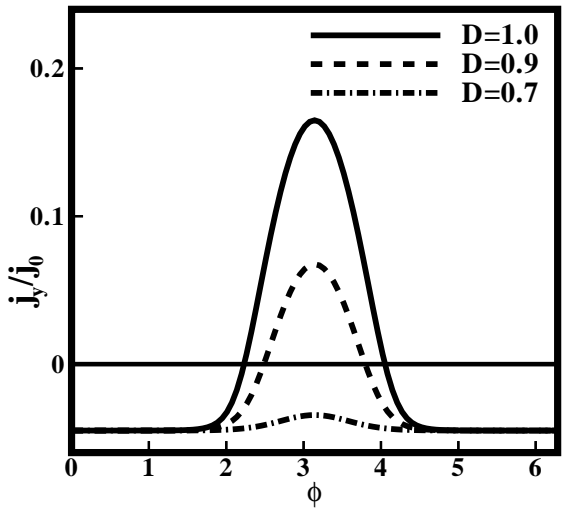}\caption{\label{fig3}Tangential
current $j_{y}$ versus phase $\protect\phi $ for $T/T_c=0.1$ and
$q=0.5$.}
\end{figure}
\begin{figure}
 \includegraphics[width=\columnwidth]{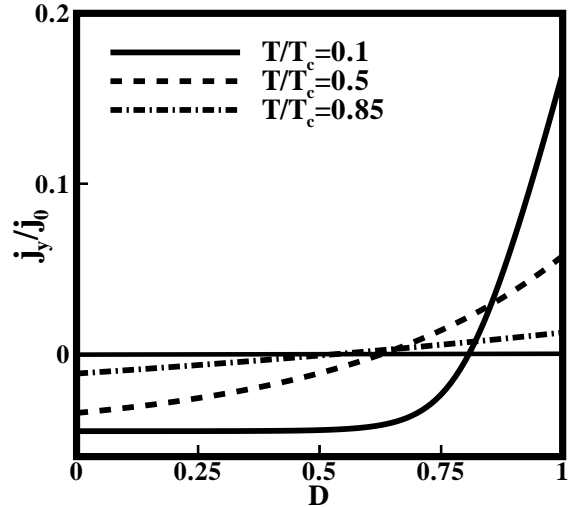}\caption{\label{fig4}Tangential
current $j_y$ versus the transparency $D$ at $\protect\phi=\pi$
and $q=0.5$.}
\end{figure}
The tangential current ${\bf j}_{y}$ depends on the order
parameters phase difference $\phi $ and is not equal to the
transport current ${\bf j}_{T}$ on the banks, in other words the
total current is not equal to the vector sum of Josephson and
transport currents. For the case ${\bf v}_{s}=0$, at the contact
the tangential current is zero and the normal component, i.e. the
Josephson current is as found for the finite transparent contact
in \cite{Zaitsev}. Detaching explicitly the Josephson current
${\bf j}_{J}$ and the spatially homogeneous (transport) current
${\bf j}_{T}$ that is produced by the superfluid velocity ${\bf
v}_{s}$, we can write the current as the sum of three terms: ${\bf
j}_{J}$, ${\bf j}_{T}$, and the ''interference'' current ${\bf
j}_{int}$. Also we have
\begin{equation}
{\bf j}_{cont}={\bf j}_{J}(\phi ,D,{\bf v}_{s})+{\bf j}_{T}( {\bf
v}_{s})+{\bf j}_{int}(\phi ,D,{\bf v}_{s}) \label{detaching}
\end{equation}
The ''interference'' current takes place in the vicinity of the
contact, where both coherent currents ${\bf j}_{J}(\phi )$ and
${\bf j}_{T}({\bf v}_{s})$ exist (see also the next section). At
first we consider the current density (\ref{j1}) for temperatures
close to the critical temperature ($T_{c}-T\ll T_{c}$). From Eqs.
(\ref{j1}) at the contact we obtain:
\begin{equation}
{\bf j}_{J}(\phi ,D,{\bf v}_{s})=\frac{1}{2}AD\sin \phi {\bf
e}_{x}
\end{equation}
\begin{equation}
{\bf j}_{T}({\bf v}_{s})=-\frac{1}{3}Ak\bf{e}_{y}, \label{j1tc}
\end{equation}
\begin{equation}
{\bf j}_{int}(\phi ,D,{\bf v}_{s})=\frac{1}{3}AkD(1-\cos \phi)
{\bf e}_{y}.\label{j2tc}
\end{equation}
where $A=\frac{1}{16}j_{0}\frac{\Delta ^{2}}{T_{c}^{2}}$,\
$k=\frac{ 14\varsigma (3)}{\pi ^{3}}\frac{v_{s}p_{F}}{T_{c}}$,
${\bf e}_{i}$ is the unit vector in the $i-$direction. This
consideration shows how the current is affected by the interplay
of Josephson and transport currents. At the contact the
''interference'' current ${\bf j}_{int}$ is anti-parallel to $
\bf{j}_{T}$ and if the phase difference $\phi =\pi $, ${\bf j}
_{int}=-2D{\bf j}_{T}$. When there is no phase difference (at
$\phi =0)$, we obtain $j_{int}=0$. So at transparency values $D$
up to ${\frac{1}{2}}$ the total tangential current at the contact
flows in the opposite direction to the transport current. Thus,
for such $D$ in the vicinity of the contact, two vortices should
exist. At arbitrary temperatures $T<T_{c}$ the current-phase
relations can be analyzed numerically. In our calculations we
define the parameter, $q$, in which $q=\frac{p_{F}v_{s}}{\Delta _{0}}$ and ${%
\Delta _{0}}=\Delta (T=0,v_{s}=0)$. The value of $q$ can be in the range $%
0<q<q_{c}$ and it's critical value $q_{c}$, corresponds to the
critical current in the homogeneous current state \cite{Bardeen}.
At $T=0$, $q_{c}=1$ and the gap $\Delta $ does not depend on $q$.
In Fig.\ref{fig2} and Fig.\ref{fig3}, we plot the Josephson and
tangential currents at the contact as functions of $\phi $ at
temperatures far from the critical (namely, $ T=0.1T_{c}$) and for
$q=0.5$ and for different values of transparency $D$. Far from
$\phi =\pi $, the tangential current is not disturbed by the
contact, it tends to its value on the bank. The Josephson
current-phase relation is the same as when the transport current
is absent. However, when $ \phi $ tends to $\pi $, for the highly
transparent contact\ $\left( D=1,0.9\right) $ the tangential
current becomes anti-parallel to the bulk current. But for $D=0.7$
the ''interference'' current is strongly suppressed and the
tangential current flows parallel to the bulk current. In
Fig.\ref{fig4}, we plot $%
j_{y}\left( D\right) =j_{T}+j_{int}$ at $\phi =\pi $ for different
temperatures. These plots show that by increasing the temperature
a counter-flow $j_{y}\left( D\right) <0$ exists in a wider
interval of transparency $D_{c}\left( T\right) <D\leq 1$ and
$D_{c}\left( T\rightarrow T_{c}\right) \rightarrow {\frac{1}{2}}.$
This numerical result coincides with analytical results
(\ref{j1tc},\ref{j2tc}).
\begin{figure}
 \includegraphics[width=\columnwidth]{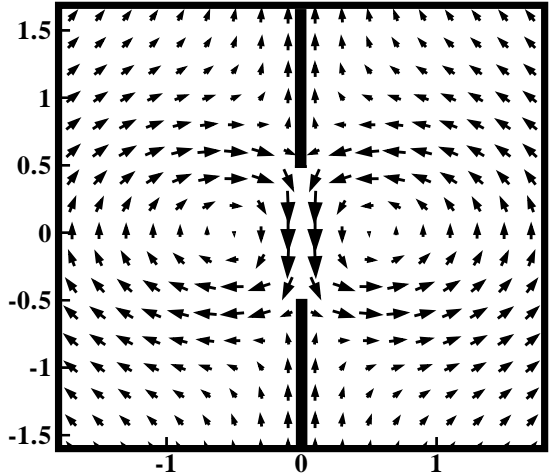}\caption{\label{fig5} Vector plot of
the current
 for $\protect\phi=\protect\pi$, $q=0.5$, $T/T_c=0.1$ and $D=0.95$.
  Axes are marked in units of the contact size $a$.}
\end{figure}
\begin{figure}
 \includegraphics[width=\columnwidth]{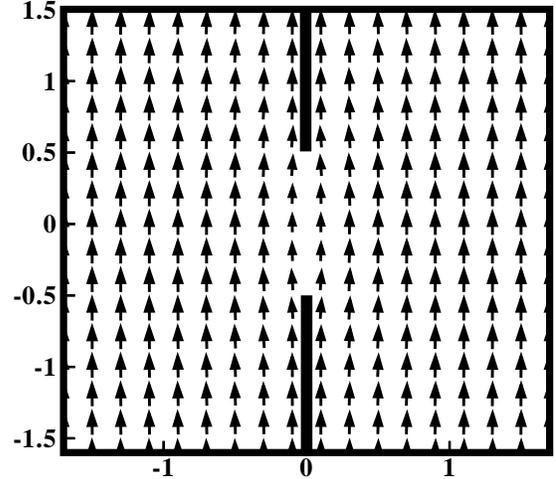}\caption{\label{fig6}Vector plot of
the current for $\protect\phi=\protect\pi$, $q=0.5$, $T/T_c=0.1$
and $D=0.7$.}
\end{figure}
\section{Spatial distribution of the current near the contact.}
In this section we consider the spatial distribution of the
current near the orifice. The superconducting current
(\ref{Current}) can be written as
\begin{equation}
{\bf j}({\bf \rho },{\bf v}_{s})=-j_{0}\frac{T}{T_{c}}
\sum\limits_{\omega >0}\left\langle \widehat{{\bf
v}}\text{Im}g({\bf  \rho },{\bf v}_{s})\right\rangle
_{\widehat{{\bf v}_{F}}}, \label{j}
\end{equation}
where, $\ j_{0}=4\pi \left| e\right| N(0)v_{F}T_{c}$. We should
note that although the current (\ref{j}) depends only on the
coordinates in the film plane, the integration over velocity
directions $\widehat{\bf v}$ is carried out over all of the Fermi
sphere as in a bulk sample. This method of calculation is correct
only for specular reflection from the film surfaces when there is
no back scattering after electron interaction with them.
 At a point, ${\bf \rho }=(x,y)$, all ballistic trajectories can
be categorized as transit and non-transit trajectories
(see,Fig.\ref{fig1}). For the transit trajectories ''1'' (their
reflected counterparts marked by ''3'' in Fig.\ref{fig1}) a
projection $\widehat{{\bf v}}_{\Vert }$ of the vector
$\widehat{\bf v}$ to the film plane belongs to the angle at which
the slit is seen from the point ${\bf \rho },$ $\widehat{{\bf
v}}_{\Vert }\in \alpha (\bf{\rho }),$ and for non-transit(marked
by ''2'' in Fig.\ref{fig1}) $\widehat{{\bf{v}}}_{\Vert }\notin
\alpha ( \bf{\rho )}$. For transit trajectories the Green's
functions satisfy boundary conditions on both banks and at the
contact. The non-transit trajectories should satisfy the specular
reflection condition (or Zaitsev's
boundary conditions (\ref{Zaitsev1})-(\ref{Zaitsev2}) for $D=0$ at $%
x=0,|y|\geq a)$. Then for the current at $T_{c}-T\ll T_{c}$ we
obtain an analytical formula
\begin{equation}
\bf{j}(\bf{\rho },\phi ,D,\bf{v}_{s})=  \label{j_T_c}
\end{equation}
$$j_{c}D\left\langle \sin \phi
\widehat{\bf{v}}\text{sgn}(v_{x})+k(1-\cos \phi
)\widehat{\bf{v}}\widehat{v}_{y}\right\rangle _{{\widehat{\bf{v}
}_{\Vert }}\in \alpha }-j_{c}k\left\langle
\widehat{\bf{v}}\widehat{v} _{y}\right\rangle
_{\widehat{\bf{v}}}$$
 where, $j_{c}\left( T,\bf{v}_{s}\right)
=\frac{\pi |e|N(0)v_{F}}{8}\frac{ \Delta ^{2}\left(
T,\bf{v}_{s}\right) }{T_{c}}$. To illustrate how the current flows
near the contact, we plot the Fig.\ref{fig5} and Fig.\ref{fig6},
for $ \phi =\pi $ and temperatures much smaller than critical
($T/T_{c}=0.1$), and for different values of transparency. At such
value of the phase $\phi $ there is no Josephson current and at
the large $D=0.95$ the current is disturbed in such a way that
there are two anti-symmetric vortices close to the orifice (see
Fig. \ref{fig5}). For the such temperature at $D=0.7$ the vortices
are absent in Fig. \ref{fig6}.
\begin{figure}
 \includegraphics[width=\columnwidth]{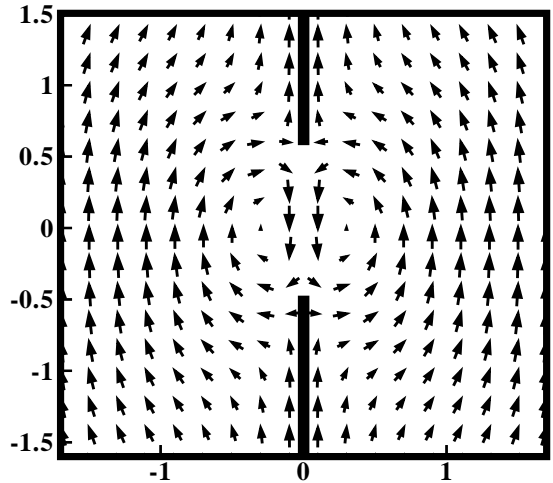}\caption{\label{fig7}Vector plot of
the current for $\protect\phi%
=\protect\pi$, $q=0.5$, $D=0.7$, and $T/T_c=0.85$.}
\end{figure}
Near the critical temperature $\left( T/T_{c}=0.85\right) $ the
vortex-like currents are restored for $D=0.7$ (see
Fig.\ref{fig7}). Far from the orifice (at the distances $l\sim \xi _{0}\gg a$%
) the Josephson current is spread out and the current is equal to
its value at infinity. Considering the current distributions and
current-phase diagrams, we observed that:\newline 1). For fixed
values of temperature and superfluid velocity, by decreasing the
transparency the vortex-like current disappears at $D=D_{c}\left(
T\right) ;$ $0.5\leq D_{c}\left( T\right) <1$\newline 2). For
intermediate values of transparency $D$ $(D_{c}\left( T\right)
<D<1)$ by increasing the temperature the vortex-like currents,
which were destroyed by the effect of electron reflection at the
contact, may be restored.\newline It is clear that both Josephson
and ''interference'' currents are the result of the quantum
interference between two coherent states. By decreasing the
transparency the interference effect will be weaker and these two
currents will decrease, while the transport current will remain
constant. On the other hand, the presence of vortices depends on
the result of competition between transport and ''interference''
current.
\begin{figure}
 \includegraphics[width=\columnwidth]{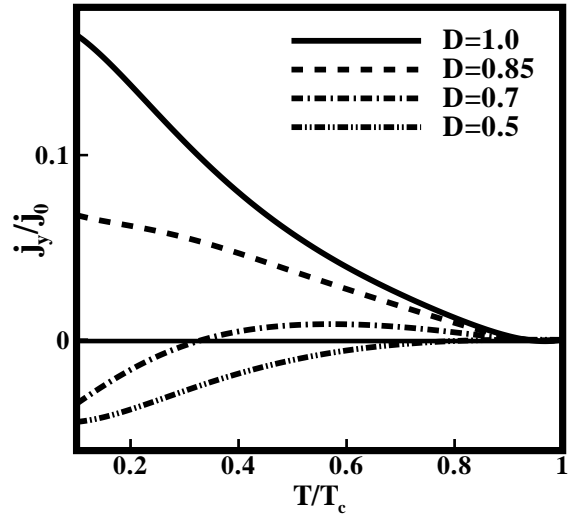}\caption{\label{fig8}Tangential
current $j_y$ versus the temperature $T$ at
$\protect\phi=\protect\pi$ and $q=0.5$.}
\end{figure}
 Thus, by decreasing the transparency the tunneling and
consequently the ''interference'' current will decrease and
vortices may be destroyed (\ref{j1tc},\ref{j2tc}). Similar to the
case $D=1$ in \cite{KOSh}, at high values of transparency, the
''interference'' current can dominate the transport current and
tangential current can be anti-parallel to the transport current,
thus the vortices appear. But for low transparency the tangential
current will be parallel to the transport current and the vortices
disappear.\\
The second point is an anomalous temperature behavior of the
effect. The vortices are the result of the coherent current
mixing. One could expect that by increasing the temperature the
vortices would disappear whereas, for intermediate values of
transparency, by increasing the temperature the vortices will be
restored. As considered in Fig.\ref{fig6} and Fig.\ref{fig7} for
the transparency $D=0.7$ the vortices at low temperature are
absent but at high temperature they are present. In the plots for
tangential current versus transparency, Fig. \ref{fig4} we can
observe this phenomenon (appearance of the counter-flow near the
contact at high temperatures).\\
Usually superconducting currents are monotonic and descendant
functions of temperature. Josephson and transport currents have
this property, but about the tangential current $j_{y}$, the
situation is totally different. At high values of transparency the
$j_{y}$ has similar behavior to the two other currents, but at low
and intermediate values of transparency at $\phi =\pi $ it has a
non-monotonic dependence on the temperature and this is the origin
of the anomalous temperature behavior of vortices. As the
temperature increases, the tangential current first increases and
then decreases. In Fig.\ref{fig8} we plotted the tangential
current ( ''interference''+ transport current) versus the
temperature for different values of transparency. We observed that
for intermediate values of transparency $0.5<D<1$, at low
temperatures and $\phi =\pi $ the tangential current has anomalous
dependence on the temperature. The reason for this dependence is
that the "interference" current flows in the opposite direction to
the transport current. This current is suppressed by the
reflection, but with increasing of the temperature it decreases
slowly than the transport current. As a consequence of that with
increasing of $T$ the tangential current can change its sign and
vortices appear. We found that for low values of transparency
$0<D<0.5$ the ''interference'' current cannot dominate the
transport current and in addition the tangential current has the
same direction as the transport current for any temperature
$T<T_{c}.$
\section{Conclusion}
We have studied theoretically the stationary Josephson effect in
the ballistic point contact with transport current on the banks in
the model S-c-S taking into account the reflection of electrons
from the contact. The contact is subject to two external factors:
the phase difference $\phi $ and the transport current tangential
to the boundary of the contact. As it was shown in \cite{KOSh}, in
the contact with direct conductivity at $\phi =\pi $ and near the
orifice the tangential current flows in the opposite direction to
the transport current, and there are two anti-symmetric
vortex-like structures. The transparency effect on the vortex-like
currents has a central role in our paper. By decreasing the
transparency $D_{c}<D<1$ the vortex-like current is destroyed. The
critical value of $D=D_{c}\left( T\right) $ depends on the
temperature $T$ and $D_{c}\left( T\rightarrow
0\right) \rightarrow 1,$ $D_{c}\left( T\rightarrow T_{c}\right) \rightarrow {%
\frac{1}{2}}$, so that we can never find a vortex for transparency
values lower than $\frac{1}{2}$ . This anomalous temperature
behavior of the vortices is the result of non-monotonic dependence
of the interference current on the temperature. The principal
possibility of the realization of the considered effect in an
experiment was described in the paper \cite{KOSh}: A
superconducting long thin-wall cylinder (with thickness of the
wall $d$ less than London penetration depth) with two cuts, such
as a distance between them is smaller than coherence length $\xi
_{0},$ is placed in magnetic field, which is parallel to the
cylinder axis. A space between the cuts plays a role of the point
contact. The phase difference $\phi $ is governed by the external
magnetic flux. The transport current $j_{T}$ flows through two
large contacts at the ends of the cylinder.
\subsection*{Acknowledgment}
We would like to thank M. Zareyan and S.N. Shevchenko for their
helpful discussions.

\end{document}